\documentstyle[12pt]{article}
\def\la{\mathrel{\mathpalette\fun <}}
\def\ga{\mathrel{\mathpalette\fun >}}
\def\fun#1#2{\lower3.6pt\vbox{\baselineskip0pt\lineskip.9pt
  \ialign{$\mathsurround=0pt#1\hfil##\hfil$\crcr#2\crcr\sim\crcr}}}

\newskip\humongous \humongous=0pt plus 1000pt minus 1000pt

\newif\ifdtup

\def\oldreffmt#1{\rlap{[#1]} \hbox to 2\parindent{}}

\def\figfmt#1{\rlap{Figure {#1}} \hbox to 1in{}}

\def\beq{\begin{equation}}
\def\eeq{\end{equation}}

\def\bq{\begin{quote}}
\def\eq{\end{quote}}
\relax

\begin{document}
\rightline{SU-ITP-900\ \ \ \ \  }
\rightline{\today}
\vskip 1.5truecm

\begin{center}

\

{\large\bf HARD ART OF THE UNIVERSE CREATION}\\
(Stochastic approach to tunneling and baby universe formation)
\vskip 2.8 cm
\centerline{{\bf Andrei Linde}\footnote{On leave of absence from: Lebedev
Physical Institute, Moscow 117924, USSR}}
\vskip 0.5cm
Department of Physics, Stanford University, Stanford CA 94305-4060,
USA\footnote{Bitnet
address LINDE@SLACVM}
\end{center}

\vskip 1.2cm
\begin{center}
{\large ABSTRACT}
\end{center}
\vskip 0.5 cm
\begin{quote}
We develop a stochastic approach to the theory of tunneling with the baby
universe formation. This method is
applied also to the theory of creation of the universe in a laboratory.

\end{quote}  \vfill

\newpage

\section{Introduction}

Few years ago many people believed that the universe is topologically
connected.
Of course, it was
possible to speculate about the universe consisting of many disconnected
pieces,
but such
speculations did not seem to have any practical consequences  since it was
assumed that
topologically  disconnected pieces of the universe cannot influence each other.
However, recently
it was understood that  topologically disconnected universes may have a
non-local interaction with
each other, which   may help us to solve the cosmological constant problem
\cite{b64}-\cite{b74}. The most popular approach to the cosmological constant
problem is based on
the possibility that even if one starts with one (topologically connected)
universe, later on many
new universes (so-called baby universes) are created from the original one due
to quantum
tunneling \cite{b20}-\cite{b23}. This idea is very interesting and it may prove
to be very
productive. However, the standard lore of the theory of tunneling with the
universe production
 is based on the Euclidean approach to quantum cosmology
\cite{b21}-\cite{b23}.   In some  cases this approach gives correct answers and
can be justified
with the help of more rigorous methods. In some other cases this approach also
gives correct
answers,  but not to the questions which were asked originally. Being applied
to
the  baby universe theory, this
method contains many ambiguities which still remain unresolved \cite{MyBook}.
One of the most difficult
problems is related to the  choice of  sign  of the Euclidean action in the
expression for the wave
function of the universe $\Psi$: Whether  $\Psi \sim \exp(-S_E)$, like the
Hartle-Hawking wave
function \cite{b55} (or  $\Psi \sim \exp(\exp(-S_E))$, with an account taken of
baby universes \cite{b24}), or $\Psi \sim \exp(S_E)$, like  the tunneling wave
function \cite{b56} ? Or maybe  (as we suspect) the true answer is much more
complicated ?

The cosmological constant problem is not the only possible application of the
theory of many
topologically disconnected universes. It is not inconceivable that only in this
context one can
really understand the origin of our own universe. Indeed, the standard idea
that
our universe
appeared from ``nothing'' (or singularity) is not much better (and conceptually
is much less clear)
than the idea that it appeared from another universe as a result of the baby
universe creation. This
brings us to another issue which was discussed recently: Is it possible to
create the universe in a
laboratory ?

In the context of the standard hot universe theory this question is not very
interesting . In order to
create a large  universe which would be as long-living as ours, one would need
to have at least as many
particles as their total number in the observable part of our universe... In
the
inflationary universe scenario
one can do the same job by producing a  scalar field $\phi$  with the potential
energy density $V(\phi)$ in a
domain of space of a radius $r$ bigger than the horizon, $r > H^{-1}(\phi) =
\sqrt{3 M^2_p\over 8\pi V(\phi)}$.
Such a domain of the inflationary universe  expands by its own laws, as a
separate universe independent of
what occurs outside it (``no-hair'' theorem for de Sitter space). The total
energy of the field $\phi$ in  a bubble
of a radius $r \sim H^{-1}$ is $E \sim {M^3_p\over\sqrt V(\phi)}$. This amounts
to  about a hundred  kilograms of
matter for the new inflationary universe scenario in grand unified theories and
to just a few Planck masses $M_p \sim
10^{-5}$ g for chaotic inflation, which may start at extremely large $V(\phi)$,
up to $V(\phi) \sim
M^4_p$ \cite{MyBook}.

This does not mean that it is easy to create an inflationary universe in a
laboratory, even with the help of chaotic inflation.   An
investigation of this problem  shows that at the classical level this process
is
forbidden;  one can
only form  small sub-critical bubbles with $r < H^{-1}$, which never expand up
to the over-critical
size $r > H^{-1}$ \cite{b100}. However, the over-critical domains can appear
due
to quantum
effects.

In particular, one may consider a small (sub-critical) bubble filled with a
scalar field with large
$V(\phi)$ and investigate a possibility that such a bubble will transform into
a
large expanding
bubble due to quantum tunneling. This possibility was considered in a series of
very interesting
papers \cite{LabTex,LabGuth} by two different methods. The results of these two
methods  agree
with each other and the authors conclude that one can actually create the
universe in a laboratory
in this way. However, there exist two problems with the methods used in
\cite{LabTex,LabGuth}.

Investigation of quantum tunneling with an account taken of gravitational
effects is  extremely
complicated. To simplify the problem, the authors of \cite{LabTex,LabGuth}
assumed that the
bubble contains vacuum in a state  $\phi = 0$,  corresponding to a local
minimum
of the effective
potential $V(\phi)$, and that the bubble  wall is very thin. The thin wall
approximation is valid in
the old inflationary universe scenario, but old inflation does not lead to a
good cosmology. This
approximation is valid also in some  inflationary models involving  several
different scalar fields
\cite{b90,EtExInf}, but the bubble formation in these models is much more
complicated, and it was
not studied in \cite{LabTex,LabGuth}. Unfortunately, we are not aware of any
situation where the
thin wall approximation would work in the standard versions of new or chaotic
inflation.

Another problem is related to the possibility to create the initial
sub-critical
bubble surrounded
by the ordinary Minkowski vacuum. In  \cite{LabTex,LabGuth} it was assumed that
this is just a
technical problem, which must have some solution. However, it is not quite
clear
whether such
solution exists at all. Indeed, the only way of producing such bubbles which we
know at present is
to heat some part of the universe up to the critical temperature $T_c$, after
which the phase
transition to the false vacuum occurs in this domain. However,  in all
realistic
models studied so
far the thermal energy density $E \sim T^4$ at the time of the phase transition
is much bigger than the  vacuum
energy density,   $T_c^4 \gg V(0)$
 \cite{MyBook}. Bubbles filled (and surrounded) by the gas of ultrarelativistic
particles behave
quite differently from the empty bubbles studied in \cite{LabTex,LabGuth}.

The problems discussed above are so complicated that it would be very desirable
to find some
simple intuitive approach which would help us to get at least partial
understanding of what is going
on. Few years ago, even before the notion of the baby universes was
introduced,
we suggested a
stochastic approach to  the inflationary universe formation  due to quantum
fluctuations in
Minkowski space and estimated the probability of this process for a large class
of models of the
scalar field $\phi$ minimally coupled to gravity \cite{b20}. Later we used a
simple generalization
of this approach to describe the probability of the universe formation in a
laboratory (at high
temperature) \cite{Trieste,Nobel}.  Stochastic approach to the universe
formation has its own
problems, to be discussed below. However, it is very simple, its results have a
clear physical
interpretation and, as we will show, it works very well being applied to  the
theory of tunneling in
field theory and quantum statistics, both in Minkowski space and in de Sitter
space. Therefore, this method may serve as a useful intermediate step towards a
more complete and rigorous investigation of the baby universe formation. In
this
paper
we will describe the stochastic approach to tunneling  and apply it to the
theory of  the baby universe
production, both in the empty Minkowski space and in a laboratory. In order to
do it we must first
say few words about Euclidean approach to tunneling and to quantum cosmology.

\section {Euclidean Methods in the Theory of Tunneling}

One of the simplest and most elegant approaches to tunneling in the scalar
field
theory \cite{Okun}  is the
Euclidean one \cite{Col}. The main idea of this approach is that tunneling is a
motion
with imaginary energy, which is equivalent to  motion in imaginary time,
i.e. in  Euclidean space. The probability of tunneling is proportional to
$\exp (-S_E)$, where $S_E$ is the Euclidean action corresponding to the
tunneling trajectory. In other words, $S_E$ is the instanton action, where the
instanton is the solution of the Euclidean field equations describing
tunneling.

A most instructive example is tunneling in a theory with the effective
potential
\begin{equation}\label{1}
V = {1\over 2}m^2\phi^2 - {1\over 4}\lambda\phi^4  + V(0) \  .
\end{equation}

The tunneling trajectories (instantons) with the minimal action possess the
O(4)
symmetry of  Euclidean space \cite{Col}. The Euclidean equation for  O(4)
symmetric tunneling is
\begin{equation}\label{2}
d^2\phi/dr^2 + (3/r)d\phi/dr = V'(\phi) \ ,
\end{equation}
with the boundary conditions $\phi(r=\infty) = \phi_o$ and
$d\phi/dr|_{r=0} = 0$. Here $r = \sqrt {x^2_i}$; the $x_i$ are the
Euclidean coordinates, i = 1,2,3,4.

In the theory (\ref{1}) with $m = 0$ equation (\ref{2}) has a class of
solutions
which are
called Fubini instantons  \cite{Fub}:
\begin{equation}\label{3}
\phi(r) = 2\sqrt {2\over \lambda} \left({\rho\over{r^2 + \rho^2}}\right) \ ,
\end{equation}
where $\rho$ is arbitrary. The corresponding Euclidean action does not depend
on $\rho$,
\begin{equation}\label{4}
S_E = 2\pi^2 \int r^3 \left({1\over 2}(d\phi/dr)^2 + V(\phi)\right) dr = {8
\pi^2\over 3\lambda} \ .
\end{equation}
The bubbles which appear after tunneling are also described by eq. (\ref{3}) if
one
understands by $r^2$ its Minkowski counterpart ${\bf r}^2 - t^2$. The
probability of the bubble formation per unit four-volume can be estimated by
the expression
\begin{equation}\label{5}
P \sim \rho^{-4} \exp {\left(-{8\pi^2\over 3\lambda}\right)}
\end{equation}
for $m = 0$.

A somewhat surprising fact is that for any nonvanishing $m$ there are {\it no}
instanton solutions in the theory (\ref{1}). This does not mean, however, that
there
is no tunneling in this theory. The point is that for any nonvanishing
$m$ the minimum of tunneling action corresponds to the $\rho \rightarrow 0$
limit of the field configurations (\ref{3}). But one can easily check that the
action
of these configurations differs from its limiting value (\ref{4}) by $\Delta
S_E
\ll
1$ for $\rho \ll {\sqrt \lambda \over m}$, i.e. for $\phi \gg m/\lambda$ in the
center
of the instanton. This means that the Fubini instantons (\ref{3})
with $\rho \ll {\sqrt \lambda \over m}$ may play the role of instantons in the
theory
(\ref{1}) with nonvanishing $m$ as well.

A similar approach can be used to study tunneling at a finite temperature.
Quantum
statistics at  finite $T$ is equivalent to quantum field theory in the
Euclidean
space
periodic in the time  direction with the period $1/T$. In the limiting case $T
\gg m$ one
should look for O(3) symmetric instantons with the time-independent field
$\phi$
and
then multiply the three-dimensional action by $1/T$  \cite{TTunn}. The
resulting
four-dimensional action for the theory (\ref{1}) is $ S \sim {19 m\over
\lambda
T}$
\cite{Witten} and the probability of tunneling is given by
\begin{equation}\label{6}
P \sim \exp\left(-  {19  m\over \lambda T}\right) \ .
\end{equation}

To describe  tunneling in the inflationary universe by Euclidean methods one
should
take into  account that the Euclidean version of de Sitter space with the
vacuum
energy
density $V(\phi)$ is  the sphere $S_4$ with the radius $H^{-1}(\phi) = \sqrt{3
M_p^2\over 8 \pi V(\phi)}$. An instanton solution in this case is a sphere
$S_4$
containing a constant scalar field $\phi $  corresponding to  an   extremum of
the
effective potential $V(\phi)$. The corresponding action  is negative and is
given by
\begin{equation}\label{7}
S = - {3 M^4_p\over 8 V(\phi)} \  .
\end{equation}
Let us
assume that the effective potential has a complicated form with a local minimum
at
$\phi  = 0$,  a local maximum at  $\phi  = \phi_1$, next local minimum at $\phi
=
\phi_2$, next local maximum at $\phi  = \phi_3$, next local minimum at $\phi  =
\phi_4$, etc. Then, according to \cite{HMoss1}, the probability of tunneling
from
$\phi = 0$ to $\phi = \phi_2$ is given by
\begin{equation}\label{8}
P = \exp\left(- S(\phi_1) + S(0)\right) = \exp\left({3M^4_p\over
8}\left(V^{-1}(\phi_1) - V^{-1}(0)
\right)\right) \ .
\end{equation}
In particular, for the theory (\ref{1}) with $m \ll \sqrt {V}/M_p$
\begin{equation}\label{9}
P =  \exp\left(- {3M^4_p m^4\over 32\lambda V^2(0)}\right) \ .
\end{equation}

Euclidean approach to tunneling  in Minkowski space at $T = 0$ can be justified
by the
standard Hamiltonian methods. Euclidean methods at $T\not= 0$ give a correct
expression for the
exponential suppression of the probability of tunneling, but some care should
be
taken when one
calculates sub-exponential terms. The situation with tunneling in de Sitter
space is much more confusing:

i) From the derivation of eq. (\ref{8})  given in \cite{HMoss1} and from all
subsequent ``proofs'' of this equation with the help of Euclidean methods by
other authors it was not clear why one should write
$S(\phi_1)$ in eq. (\ref{8}) rather than, say, $S(\phi_2)$ or $S(\phi_3)$.
Indeed, the instantons discussed in \cite{HMoss1} describe de Sitter space
filled by a homogeneous scalar field which may correspond  to {\it any}
extremum
of $V(\phi)$.

ii) According to its derivation,  eq. (\ref{8}) should work equally well for $H
> m$ and for $H < m$ since the
instantons used for its derivation exist independently of the value of $m$.
However, stochastic methods to be used in the next section show that eq.
(\ref{8}) is valid only if
$V^{\prime\prime}(\phi) \gg H^2$ all the way from $\phi_0$ to $\phi_1$.

iii) The instanton solution obtained in \cite{HMoss1} describes a homogeneous
field $\phi$ rather than a
 bubble.   Therefore   eq. (\ref{8}) was interpreted originally as an
expression
for the probability of
tunneling which occurred simultaneously in  the whole universe. This
interpretation proved to be incorrect
\cite{b61}. Another interpretation   \cite{HMoss2} was that eq. (\ref{8})
described appearance of large bubbles with an
interior which looked homogeneous on a scale $l > H^{-1}$. Unfortunately, it
remained unclear
how   solutions with a constant field $\phi$ could describe the process of an
inhomogeneous tunneling with bubble formation.

It took few years to prove that eq.
(\ref{8}) and its interpretation in \cite{HMoss2} were actually correct,  but
only under some conditions which did not follow from its Euclidean derivation.
The proof was given within the stochastic approach to tunneling \cite{b60,b61},
which we will  discuss now.

\section{Stochastic Approach to Tunneling}

The main idea of the stochastic approach  can be illustrated  on an example of
tunneling in the theory (\ref{1}) with $m = 0$ in Minkowski space. Equation of
motion for the bubble in
Minkowski space is
\begin{equation}\label{10}
\ddot\phi = d^2\phi/dr^2 + (2/r)d\phi/dr  - V'(\phi) \  .
\end{equation}
At the moment of its formation, the bubble wall does not move. Then it starts
growing if
$\ddot\phi > 0$, which requires that
\begin{equation}\label{11}
 |d^2\phi/dr^2 + (2/r)d\phi/dr| < - V'(\phi) \ .
\end{equation}
A bubble of a classical field is formed only if  it
contains a sufficiently big field $\phi$ (it should be over the barrier, so
that
$dV/d\phi < 0$) and if the bubble itself is sufficiently large. If the size of
the bubble
is too small, the gradient terms are bigger than the term $|V'(\phi)|$, and the
field
$\phi$ inside the bubble does not grow. Typically, the second term in
(\ref{11})
somewhat compensates the first one. To make a very rough estimate, one may
write
the
condition (\ref{11}) in the form
\begin{equation}\label{12}
{1\over 2} r^{-2} \sim
{1\over 2} k^2 <  {1\over 2} k^2_{max}  \sim \phi^{-1} |V'| =  \lambda\phi^2 =
{1\over 3} m^2(\phi) \equiv
 {1\over 3} V''(\phi) \ .
\end{equation}
Let us estimate the probability of an event when vacuum fluctuations
occasionally build up a configuration of the field satisfying this condition.
In order to do it one should remember that the dispersion of quantum
fluctuations
of the  field $\phi$ with $k < k_{max}$ is given by
\begin{equation}\label{13}
<\phi^2>_{k<k_{max}} = {1\over 4\pi^2}\int_{0}^{k_{max}}{k^2 dk\over
\sqrt {k^2 + m^2}} \ .
\end{equation}
For the massless field  $k^2_{max} = 2C^2 \lambda\phi^2$, where  the factor
$C^2 =
O(1)$  reflects some uncertainty in our estimate of the value of $k^2_{max}$.
This gives
\begin{equation}\label{14}
<\phi^2>_{k<k_{max}} =    {1\over 4\pi^2}\int_{0}^{k_{max}}{k dk}  =
{k^2_{max}\over 8\pi^2} = {C^2 \lambda\phi^2\over
8\pi^2} \ .
\end{equation}

This is an estimate of the dispersion of perturbations which may sum up
to produce a  field $\phi$ which satisfies the condition (\ref{12}).  Of
course,
this
estimate is rather crude.  But
let us  nevertheless use eq. (\ref{14}) to evaluate the probability that these
fluctuations build up a bubble of the field $\phi$ of a radius $r >
k^{-1}_{max}$.
This can be done with the help of the Gaussian distribution:
\begin{equation}\label{15}
P(\phi) \sim \exp(-{\phi^2\over
2<\phi^2>_{k<k_{max}}}) =   \exp (-{2\pi^2\over C^2 \lambda}) \ .
\end{equation}
Note that the factor in the exponent in (\ref{15}) to within a factor of $C^2 =
3/4$
coincides with the Euclidean action $S_E$ in eq. (\ref{4}). Taking into account
the very
rough method we used to calculate the dispersion of the perturbations
responsible for
tunneling, the coincidence is rather impressive.

This method  also helps to study tunneling in
the theory (\ref{1}) with $m \not= 0$, in which the instanton solutions do not
exist.
Indeed, a generalization of the previous calculation  for this case gives the
following
result for the tunneling probability:
\begin{equation}\label{16}
P(\phi) \sim  =   \exp \left(-{2\pi^2\over C^2  \lambda}\left(1 +
O({m^2\over\lambda\phi^2})\right) \right) \ .
\end{equation}

This means that tunneling is possible for $m \not= 0$. There is an additional
exponential suppression of the probability of tunneling in this theory as
compared
with (\ref{15}). However, this additional suppression disappears if tunneling
occurs by
formation of bubbles with $\phi > m/\lambda$. This is in an agreement with our
arguments given in the previous section. For a discussion of other aspects of
tunneling
in the theory (\ref{1}) within the stochastic approach see also
\cite{Arn,ELSHER}.

The agreement between the results of the  method discussed
above and a more complicated Euclidean approach becomes even more  impressive
if
one
remembers that most of the results obtained in the tunneling theory by
Euclidean
methods \cite{MyBook} can  easily be reproduced (with an accuracy of the
coefficients
of $O(1)$ in the exponent) by this  simple method.

For example, let us consider  the theory (\ref{1}) at a temperature $T \gg m$.
In this
case
\begin{equation}\label{17}
<\phi^2>_{k<k_{max}} = {1\over 2\pi^2}\int_{0}^{k_{max}}{k^2 dk\over
\sqrt {k^2 + m^2}\left(\exp{\sqrt{k^2 + m^2(\phi)}\over T} - 1 \right)}  \sim
{T
\over
2\pi^2}\int_{0}^{k_{max}}{k^2 dk\over {k^2 + m^2}} \ .
\end{equation}
Note that the field inside the bubble should be somewhat bigger than  $2
m/\sqrt{\lambda}$ since otherwise $V(\phi) > V(0)$ and the phase transition is
energetically impossible. At  $\phi > 2 m/\sqrt{\lambda}$ the main contribution
to
$V'$ is given by $- \lambda \phi^3$  and the expression for $k_{max}$ is the
same as
in the previous example, $k^2_{max} = 2C^2 \lambda\phi^2 > m^2$. This gives
\begin{equation}\label{18}
P(\phi) \sim   \exp \left(-{\pi^2 \phi\over C T \sqrt{2
\lambda}} \right) \ .
\end{equation}
The probability increases with a decrease of $\phi$,
but $\phi$ should remain bigger than $2m/\sqrt{\lambda}$. To make a final
estimate, let us
take $\phi \sim 3 m/\sqrt{\lambda}$. The result is
\begin{equation}\label{19} P(\phi)
\sim   \exp \left(-{3 \pi^2 m \over C T \sqrt{2}  \lambda} \right) \ .
\end{equation}
This coincides with the result of the Euclidean approach (\ref{6}) for $C = {3
\pi^2
\over19 \sqrt {2}} \sim  1.1$.

Similar methods can be applied to tunneling in the inflationary universe. There
are
some differences though. First of all, the total energy of the bubble grows
exponentially during inflation. In other words, the total energy of the scalar
field
inside the bubble is not conserved and there is no restriction $V(\phi) > V(0)$
mentioned above. Moreover, there is no need for the new phase to be
energetically
favorable for the exponentially large bubbles of the new phase to appear. It is
sufficient for the radius of the bubble to be bigger than $H^{-1}$ and for the
{\it
homogeneous} field $\phi$ inside it to be stable with respect to
rolling back to $\phi = 0$. In other words, one should not worry about the
bubble walls
moving towards the center of the bubble until it disappears. Indeed,
if the bubble walls originally are displaced at a distance $r > H^{-1}$ from
the
center of
the bubble, this distance later grows with a speed greater than the speed of
light due to
the exponential expansion of the universe. Therefore, the local motion of the
walls
towards the center cannot lead to a disappearance of the bubble with an initial
size $r > H^{-1}$. This means that for tunneling to occur it is sufficient if
the bubble
of a radius $r > H^{-1}$ is formed containing any field $\phi > \phi_1$, where
$\phi_1$ corresponds to the nearby maximum of $V(\phi)$.

Calculation of $<\phi^2>_{k<H}$ during inflation is  rather complicated.
Fortunately, for
 the theory with the effective potential $V(\phi) = {m^2\over 2}\phi^2 + V(0)$
the
result is well known \cite{BD,b99},
 \begin{equation}\label{20}
<\phi^2>_{k<H} = {3 H^4\over 8\pi^2 m^2} \ ,
\end{equation}
where $H^2 = {8 \pi V(0)\over 3 M^2_p}$.
According to (\ref{15}), this gives the following estimate of the probability
of
formation of a
bubble of a field $\phi \ge \phi_1 = m/\sqrt\lambda$ of a radius $r > H^{-1}$
(i.e. of the
probability of tunneling):
\begin{equation}\label{21}  P =  \exp\left(- {3M^4_p m^4\over 16
\lambda V^2(0)}\right) \ .
\end{equation}
This agrees, up to the coefficient 2 in the exponent, with the Euclidean result
(\ref{9}).  The
agreement becomes complete if one investigates the probability distribution in
a
more detailed
way, without approximating $V(\phi)$ by its quadratic part.

The wave-lengths of all vacuum fluctuations of the scalar field $\phi$ grow
exponentially
in the expanding universe. When the wavelength of any particular fluctuation
becomes greater than
$H^{-1}$, this fluctuation stops propagating, and its amplitude freezes at some
nonzero value
$\delta\phi (x)$ because of the large friction term $3H\dot{\phi}$ in the
equation of motion of the
field $\phi$. The amplitude of this fluctuation then remains almost unchanged
for a very long
time, whereas its wavelength grows exponentially. Therefore, the appearance of
such a frozen
fluctuation is equivalent to the appearance of a classical field $\delta\phi
(x)$ that does not
vanish after averaging over macroscopic intervals of space and time. The
average
amplitude of the
frozen field generated during a typical time $H^{-1}$ is given by $\delta\phi
(x) = H/2\pi$ \cite{b99}. This
field looks constant on a scale $H^{-1}$, but on a bigger scale it is
inhomogeneous. If one is
interested in the value of this field in each particular point, one should take
into account that
each new wave is frozen with a different phase. As a result, the  field $\phi$
in each particular
point moves as a   Brownian particle. This makes it possible to write a
diffusion equation for the
probability distribution to find a field $\phi$ with the wavelength bigger than
$H^{-1}$ in a given
point at a given time \cite{b60}:
\begin{equation}\label{22}
\frac{\partial P}{\partial t} = \frac{\partial}{\partial \phi}
\left(\frac{\partial({\cal D}P)}{\partial \phi} + \frac{P}{3H} \frac{\partial
V}{\partial
\phi}\right) \ .
\end{equation}
Here the coefficient of diffusion ${\cal D}$ is given by $H^3/8\pi^2$.
If the probability of tunneling is sufficiently small, then  the distribution
$P$ with a good
accuracy is given by the stationary solution of eq. (\ref{22}),
\begin{equation}\label{23}
P \sim \exp\left({3M^4_p\over 8 V(\phi)}\right) \ ,
\end{equation}
To describe tunneling from the point $\phi = 0$ over the barrier at $\phi_1$
one
should consider
the probability  of formation of a bubble of a field $\phi \ge \phi_1 $ of a
radius $r >
H^{-1}$,
\begin{equation}\label{24}
P \sim \exp\left({3M^4_p\over 8 }\left(V^{-1}(\phi_1) - V^{-1}(0)\right)\right)
\ ,
\end{equation}
where the last term is necessary for  normalization of the probability
distribution. This
coincides with the Euclidean result (\ref{8}). However, now we have a clear
interpretation of this
result. (For a more detailed discussion see \cite{b60,b61,MyBook}.) We can also
understand the
limits of its validity. Namely, eq. (\ref{22}) is valid only during inflation
and only if
$V^{\prime\prime} \ll H^2$ in the interval from $\phi = 0$ to $\phi = \phi_1$.
This condition does not
follow at all from the derivation  of eq. (\ref{8}) in the Euclidean approach.

This comment proves to be very important in the context of quantum cosmology.
Indeed,
the stationary solution  (\ref{23}) looks as a square of the Hartle-Hawking
wave
function of the
ground state of the universe \cite{b55} obtained by Euclidean methods.
Therefore one  could consider eq. (\ref{23}) as a
confirmation of the validity of the Hartle-Hawking wave function, which was
criticized in
\cite{b56}. However, in all realistic theories the condition $V^{\prime\prime}
\ll H^2$ of validity
of eqs.  (\ref{22}), (\ref{23}) is not satisfied near the absolute minimum of
$V(\phi)$. As a result,
all solutions of eq. (\ref{22}) in realistic theories are non-stationary. The
probability distribution   (\ref{23}) can be used
at intermediate stages of inflation for an approximate description of
tunneling,
when the probability distribution is quasi-stationary, but it does not have any
fundamental significance and cannot be used to describe the universe as a
whole.

An important physical consequence of the Brownian motion of the field $\phi$
during inflation is
the self-reproduction of the inflationary universe. This process is especially
interesting in the context of chaotic inflation  \cite{b19,MyBook}. It proves
that at large $\phi$ the Brownian motion
is much more rapid than the classical rolling of the field to the minimum of
its
potential energy
density. As a result, many inflationary domains are formed  which contain {\it
growing} field
$\phi$. These domains expand much faster than the domains with small $\phi$.
This leads to a
paradoxical situation where the main part of the physical volume of the
universe
(in the
synchronous coordinate system) becomes occupied not by the field corresponding
to the minimum
of $V(\phi)$, but by the field fluctuating at a density close to the Planck
energy density $M_p^4$
\cite{b19,MyBook}. It is quite
clear that if the process of the baby universe formation can occur at all,
inflationary domains with
$V(\phi) \sim M^4_p$ are the best place for it. We will return to this question
in the end of the
paper.

\section{Stochastic Approach to the Baby Universe Formation}

Now that we demonstrated usefulness and reliability of the stochastic approach
in many
different situations where it can be confirmed by other methods, we will take a
deep breath and
try to apply it  to the investigation of the possibility of creating an
inflationary universe from Minkowski space
\cite{b20,Trieste,Nobel}.    The issue here is that  quantum fluctuations in
Minkowski space can
bring into being an inflationary domain of a radius $r > H^{-1}(\phi)$, where
$\phi$ is a scalar
field produced by quantum fluctuations in this domain.  The ``no-hair'' theorem
for de Sitter space
implies that such a domain inflates in an entirely self-contained manner,
independent of what
occurs in the surrounding space.  We could then conceive of a ceaseless process
of creation of
inflationary mini-universes that could take place even at the very latest
stages
of development of
the part of the universe that surrounds us.

	Without pretending to provide a complete description of such a process,
let us attempt to estimate its probability in theories with
 $V(\phi) = {{\lambda \phi^n}\over {n {M_p}^{n-4}}}$ using the methods
elaborated in the previous section.  A
domain formed with a large field $\phi$ can behave as a part of inflationary
universe
only if $\phi \ga M_p$ and the gradient and kinetic energy of this field
${1/over 2}{(\partial_\mu\phi)}^2$ is smaller than  $V(\phi)$ in its interior.
The last condition implies
that the size of the domain must exceed $r \sim \phi {V^{-1/2}(\phi)}$.  Such a
domain could arise through the
build-up of quantum fluctuations $\delta\phi$ with a wavelength
\begin{equation}\label{eq100}
r \sim k^{-1} \geq  k^{-1}_{max} \sim \phi {V^{-1/2} (\phi)}\sim {m^{-1}(\phi)}
\ .
\end{equation}
(Note that $m^{-1}(\phi) > H^{-1}(\phi)$ during inflation.) One can estimate
the
dispersion $<\phi^2>_{k<m}$
of such fluctuations using the simple
formula
\begin{eqnarray}\label{eq101}
<\phi^2>_{k<m}  \sim
{{1\over{4\pi^2}}\int_0^{k_{max}}	{k dk}}
\sim {m^2\over\pi^2} \sim {V(\phi)\over{8 \pi^2\phi^2}} \ ,
\end{eqnarray}
and for a Gaussian distribution  $P(\phi)$ for the appearance of a field
$\phi$
which is
sufficiently homogeneous on a scale $r > m^{-1}(\phi)$, one has \cite{b20}
\begin{equation}\label{eq102}
P(\phi) \sim \exp(-{\phi^2 \over 2 <\phi^2>_{k<m}}) \sim
\exp({-C{{\pi^2\phi^4}\over V(\phi)}}) \ ,
\end{equation}
where C = O(1).  In particular, for a theory with $V(\phi) = {\lambda\over 4}
\phi^4$,
\begin{equation}\label{eq103}
P(\phi) \sim \exp(-C{{4\pi^2}\over \lambda})  \ .
\end{equation}
Note, that eq. (\ref{eq103}) does not show any suppression of the probability
of
the baby universe formation
due to smallness of gravitational effects; $P(\phi)$ may be quite large in the
theory with a large coupling
constant $\lambda$.\footnote{One should remember that in a theory of several
scalar fields with different
coupling constants the probability of creation of an inflationary universe is
given by the {\it largest} coupling
constant, whereas the density perturbations $\delta\rho/\rho$ is determined by
the {\it smallest}  coupling
constant $\lambda$ \cite{MyBook}.} Some suppression may appear at the
sub-exponential level, since the
``phase space'' of all possible inflationary universes in this theory is
constrained by the condition $\phi >
M_p$.

The probability of the baby universe formation in the theory of a massive
scalar
field can be estimated by
\begin{equation}\label{eq104}
P(\phi) \sim \exp(-C{{8\pi^2\phi^2}\over m^2}) \  .
\end{equation}
The leading contribution is given by the universes with the smallest $\phi$
compatible with inflation in this
model, $\phi \sim M_p$:
\begin{equation}\label{eq105}
P(\phi) \sim \exp(-C{{8\pi^2M^2_p}\over m^2}) \  .
\end{equation}
This means that the baby universe formation should be very efficient in the
theories in which the heaviest
scalar particles have masses comparable with $M_p$.

	The main objection to the possibility of quantum creation of an inflationary
universe in Minkowski space is that the energy conservation forbids the
production of an
object with positive energy out of  vacuum with vanishing
energy density.\footnote{This problem will not appear  when we will study the
process of the universe
formation at large temperature.} Within the scope of the classical field
theory,
in which the energy density is
everywhere positive, such a process would therefore be impossible.  But at the
quantum level, the energy
density of the vacuum is zero by virtue of the cancellation between the
positive
energy density of classical
scalar fields, along with their quantum fluctuations, and the negative energy
density associated with quantum
fluctuations of fermions, or the bare negative energy of the vacuum.  The
creation of a positive
energy-density domain through the build-up of long-wave fluctuations of the
field
$\phi$ is inevitably accompanied by formation of a region surrounding that
domain in
which the long-wave fluctuations of the field $\phi$ are suppressed, and the
vacuum
energy density is consequently negative.  Here we are dealing with the familiar
quantum fluctuations of the vacuum energy density about its zero point.

 It is important  that
from the point of view of an external observer, the total energy of the
inflationary
region of the universe (and indeed the total energy of the closed inflationary
universe)
does not grow exponentially;  the region that emerges forms a universe distinct
from
ours, to which it is joined only by a connecting throat (wormhole).
The shortfall of long-wave fluctuations of the field $\phi$ surrounding the
inflationary
domain is quickly replenished by fluctuations arriving from neighboring
regions,
so
the negative energy of the region near the throat can be be rapidly spread over
a large
volume around the inflationary domain. In such a scenario the total energy of
the
inflationary domain plus the energy of vacuum surrounding it will remain zero,
but
after the negative energy  of vacuum surrounding the inflationary domain will
be
distributed all over the rest of the universe an observer near the inflationary
domain
would see it as an evaporating black hole. To show that this is a viable
possibility one can make a naive
estimate of the
Schwarzschild radius of the inflationary domain from the point of view of an
external
observer. The total energy of matter inside this domain is $E \sim V(\phi)
m^{-3}(\phi)$. If one neglects the
gravitational defect of mass (this is the place where the estimate is naive),
then the  Schwarzschild mass of
this domain is equal to $E$ and the Schwarzschild radius is $E/M^2_p \sim
m^{-1} \cdot H^2/m^2 \gg m^{-1}$. Thus, the
Schwarzschild radius of the inflationary domain is much bigger than the  size
of
the domain $ r \sim m^{-1}$,
so it really looks like a black hole. One can  show also that the time of
evaporation of such a black hole is
microscopically small, but it is much bigger than $m^{-1}(\phi)$. For example,
it can be shown that in the
theory ${{\lambda}\over 4}\phi^4$ the time of the black hole evaporation is of
the order
\begin{equation}\label{a}
t \sim {{\phi^3}\over {{{\lambda}^{3/2}} M_p^4}} > {(\lambda
m(\phi))}^{-1} \ .
\end{equation}
 This means that for a very distant observer all what happens will look as a
kind of an unusual
long-living quantum fluctuation, an observer at a not too big distance from the
inflationary domain will see it as an evaporating black hole surrounded by
space
with
negative vacuum energy, whereas an observer inside the inflationary domain
would
believe that he lives inside an inflationary universe.

Since the universes created in the empty space do not carry any momentum, the
process of their
formation cannot be localized at any particular point. Their main role in the
theory is the same as the role of
the baby universes considered in \cite{b24} -- \cite{b31}: They modify the
properties of the vacuum state.
However, the methods discussed in this section can be easily extended for the
investigation of the process of
formation of  inflationary universes in a laboratory, where this process can be
localized.  For example, one can consider the process of production
of inflationary bubbles inside a  domain containing matter heated up to a
temperature T. To this end one
should just replace expression (\ref{eq101}) for  $<\phi^2>_{k<k_{max}}$ in eq.
(\ref{eq102})  by its  counterpart
calculated at a  temperature $T \gg m(\phi) \sim V^{1/2}/\phi$:
\begin{equation}\label{bb}
<\phi^2>_{k<k_{max}}   \sim {TV^{1/2}(\phi)\over {2 \pi^2 \phi}} \ .
\end{equation}
Note, that this expression for $T \gg  V^{1/2}/\phi$ is much larger than its
empty space predecessor (\ref{eq101}). This means (not unexpectedly) that
heating  leads to a more efficient creation of universes with a given $\phi$,
and this process is localized in the part of the universe with a large
temperature (i.e. in a laboratory). The corresponding probability is given by
 \begin{equation}\label{b}
P_T(\phi) \sim \exp\left(-{\phi^2 \over 2 <\phi^2>_{k<m}}\right)
\sim \exp\left({-C{{\pi^2\phi^3}\over{TV^{1/2}(\phi)}}}\right) \ .
\end{equation}

For the theory ${{\lambda}\over 4}\phi^4$ the maximum contribution is given by
$\phi \sim M_p$,
\begin{equation}\label{c}
P_T \sim \exp\left({-C{{\pi^2M_p}\over{T \sqrt\lambda}}}\right) \  ,
\end{equation}
whereas for the theory ${m^2\phi^2\over 2}$ the result is
\begin{equation}\label{d}
P_T \sim \exp\left({-C{{\pi^2M^2_p}\over{T m}}}\right) \  .
\end{equation}
These expressions remain exponentially small at $T \ll M_p$, for all values of
$\lambda$ and $m$.  Moreover,
 even if one could rise the temperature up to $M_p$, one would not  produce
inflationary universes that way.
Eqs. (\ref{b})-(\ref{d})  are valid only if $T^4 \la V(\phi)$, since otherwise
the energy density of hot matter
inside the bubble is  bigger than the effective potential, and inflationary
regime cannot be realized. (This is
the same problem as the one discussed in the Introduction in relation to papers
\cite{LabTex,LabGuth}.)
Therefore, the probability of creation of inflationary universe due to
high-temperature effects in the theory ${{\lambda}\over 4}\phi^4$ is
bounded from above by
\begin{equation}\label{e}
P_{max}(\phi) \sim
\exp\left({-C{{\pi^2\phi^3}\over{V^{3/4}(\phi)}}}\right) \sim
\exp\left({-C{{\pi^2}\over{\lambda^{3/4}}}}\right) \ .
\end{equation}
This maximum is reached at $T \sim
\lambda^{1/4} M_p$. The corresponding expression in the theory ${m^2\phi^2\over
2}$ is
\begin{equation}\label{f}
P_{max}(\phi) \sim  \exp\left({-C{{\pi^2}{\left(M_p\over
m\right)}^{3/2}}}\right) \ ,
\end{equation}
This maximum is achieved at $T \sim \sqrt{m M_p}$.

One could hope that it might be easier to create the universe in the new
inflationary scenario, since inflation
occurs there on a much smaller energy scale. Unfortunately, the result is quite
opposite: The smaller is the
energy density, the larger is $H^{-1}$, the larger is the size of the domain to
be produced, the smaller is the
probability of such event. As an example, we will consider here the theory with
the effective potential often
used in the new inflationary universe scenario,
\begin{equation}\label{g} V(\phi) =
{\lambda\phi^4}\left(\log{\phi\over\phi_0} - {1\over 4}\right)+ V(0)  \  .
\end{equation}
Here $V(0) =
{\lambda\phi_0^4/ 4}$, $\phi_0$ is the position of the minimum of $V(\phi)$,
the
mass of the field in this
minimum is equal to $m = 2\sqrt \lambda \phi_0$. Inflationary domain should
contain the field $\phi \ll
\phi_0$. This means that  the deviation from the minimum due to quantum
fluctuations is given by $\phi_0$.
The  size of the domain should exceed $H^{-1} = {M_p\over \phi_0^2}\sqrt{3\over
2\pi\lambda}$. It should be even larger
for tunneling to small $\phi$, with $V^{\prime\prime} \ll H^2$, but here we
will
consider the simplest case
when $V^{\prime\prime}$ is just few times smaller than $H^2$. Note, that $m \gg
H$ for $\phi_o \ll M_p$. In
this case
\begin{equation}\label{h}
<\phi^2>_{k<H} \  \sim	 {{1\over{4\pi^2}}\int_0^{H} {{k^2
dk}\over{\sqrt{k^2+m^2}}}} \sim {H^3\over 12 \pi^2 m} \ .
\end{equation}
This gives the following estimate of the probability of the baby universe
creation  in this theory:
\begin{equation}\label{i}
P   \sim
\exp(-{\phi_0^2 \over 2 <\phi^2>_{k<H}})  \sim  \exp\left(- C{{4 \pi^2}\over
\lambda} \cdot
\left({M^p\over\phi_0}\right)^3\right) \ .
\end{equation}
 This is much smaller than the corresponding probability in the chaotic
inflation scenario (\ref{eq103}).

By a similar method one can get the following estimate of the probability to
create the universe in a laboratory
at $T \gg m$:
\begin{equation}\label{j}
P _T    \sim  \exp\left(- C{{2 \pi^2}\phi_0\over T\sqrt\lambda} \cdot
\left({M_p\over\phi_0}\right)^3\right) \  .
\end{equation}
{}From the condition $T^4 <  V(0) = \lambda\phi_0^4$ it follows that
\begin{equation}\label{k}
P _T    <  \exp\left(-C{{\pi^2}\over \lambda^{3/4}} \cdot
\left({M_p\over\phi_0}\right)^3\right) \  .
\end{equation}
Again, for $\phi_0 \ll M_p$ this probability is much smaller than that in the
chaotic inflation scenario, see
eq. (\ref{e}).

Thus, we see that the stochastic methods, which proved to be very efficient in
the theory of tunneling,
may help us  to understand such issues as the baby universe formation in the
vacuum and at a finite
temperature. However,  several problems are to be resolved  before one gets too
excited.

First of all, our estimates give us the probability of formation of a large
domain with the energy density
dominated by the potential energy density of the slowly changing  field $\phi$.
Such a domain should behave as a part of
de Sitter space.  However, if it can be considered as a part of a {\it closed}
 de  Sitter space, then in the beginning it may be rapidly contracting.
Contraction may lead to appearance of large
gradients of the field $\phi$ and to the collapse of the domain, instead of its
exponential expansion.

Even though this problem is very complicated, we believe that it does not cause
any difficulties in the case when
the size of the bubble is much bigger than $H^{-1}$. Indeed, the acceleration
of
test particles in de Sitter space
is always large and {\it positive}, $\ddot r/r = H^2$. One may expect that the
bubble walls at the moment of its
formation,  being formed  by overlapping of long-wave quantum fluctuations in
Minkowski space, cannot move
towards the center of the bubble  with the speed bigger than the speed of
light.\footnote{In the standard
tunneling theory in Minkowski space the initial speed of the bubble walls is
equal to zero, whereas
the bubbles formed in the inflationary universe may expand (but not contract)
from the very beginning.}  If this is true,
then  the bubble of de Sitter space of a size exceeding $H^{-1}$ becomes
exponentially expanding within a
fraction of the Hubble time $H^{-1}$ due to the acceleration $\ddot r/r = H^2$
mentioned above.

The second problem can be explained as follows. When we studied  formation of
an
inflationary bubble, we did
not take into account the backreaction  of the changing   metric inside the
bubble on the spectrum of vacuum
fluctuations during the process of the bubble formation. This is a good
approximation for small bubbles with a radius
$r < H^{-1}$, but for  large bubbles the situation is not that simple. Let us
try to visualize the
process of growth of the field $\phi$ in Minkowski space due to overlapping of
many
 long-wave fluctuations. The wavelength of each such wave should exceed
$O(m^{-1}(\phi))$. Therefore one may expect that the total time necessary for
the maxima of these waves to
overlap  should be also of the order of $m^{-1}(\phi) \gg H^{-1}$. If the
bubble
will start expanding
exponentially at any stage of this process,   then its expansion during the
time $O(m^{-1}) \gg H^{-1}$  may  push away all new incoming long-wave
fluctuations of the scalar field which could
lead to  further growth of the field inside the bubble. This may make the
process of direct formation of inflationary baby universes with
$m^{-1}(\phi) \gg H^{-1}$  by the mechanism suggested above either strongly
suppressed or even entirely
forbidden.   In the last case, the estimates we made could be only used to
calculate the probability of formation of
small, sub-critical bubbles with the energy density dominated by $V(\phi)$. (As
we  emphasized in the
Introduction, this is an important problem by itself.) Then it will be
necessary
to consider  further tunneling of
a sub-critical bubble to a  large exponentially expanding bubble along the
lines
of \cite{LabTex,LabGuth}, but
without the use of the thin-wall approximation.  One may expect that the
probability of such a double tunneling
will be even smaller than the probability of each of such events separately.
In
such case one may use our
estimates as  an  upper bound  on the probability of the baby universe
formation.

However, this problem should not be exagerrated.  First of all, this problem
may
not appear at all for the
bubbles containing the field $\phi \sim M_p$ with $m^{-1}(\phi)$ being just few
times larger than $H^{-1}$.
This is quite sufficient for a production of the universe of our type with a
size $l > 10^{28}$ cm \cite{MyBook}. Moreover,
inflation of the interior of the bubble can push away incoming long-wave
perturbations {\it only if inflation
there already started}. But in this case inflation itself produces long-wave
perturbations by stretching the
short-wave ones, which always exist in the interior of the bubble. This is the
same mechanism which leads to
appearance of extremely large long-wave perturbations in the inflationary
universe, eq. (\ref{20}). Indeed,
it is known that the total contribution to $< \phi^2 >$  in de Sitter space
from
perturbations with the
wavelength smaller than $H^{-1}$ is the same as in Minkowski space, and the
contribution of long-wave
perturbations with the wavelength larger than $H^{-1}$ is even greater than in
Minkowski space, due to the
process of stretching of the wavelengths mentioned above. This suggests that
our estimate for the
probability of bubble formation with $\phi \sim M_p$, $m^{-1}(\phi) \sim
H^{-1}$
has a good chance to be correct, whereas the
probability of the universe formation with $m^{-1}(\phi) \sim H^{-1}$ may be
even somewhat higher than we
expected.

\section{Discussion}

 It is quite clear that the estimates which were obtained by the stochastic
approach to tunneling cannot serve
as a substitute of a  complete investigation by more regular methods;  a much
more detailed investigation is
needed to prove that the process of the baby universe formation can actually
occur either in the empty
Minkowski space or in a laboratory.
 However, the use of Euclidean methods in  quantum
cosmology never have  been justified and often give ambiguous results, whereas
the Hamiltonian approach
usually is extremely complicated and may have its own conceptual limitations
being applied to a system of
many universes with different ``times''. Therefore, it would be very useful in
the beginning to get at least
partial understanding of the processes we are going to study. This was the main
goal of the present
investigation.

Our estimates indicate that the process of the baby universe formation in
Minkowski space is actually possible.
Moreover, in the theories with large coupling constants or with scalar
particles
with masses of the order of
$M_p$ this process may be even quite probable.  This process by itself does not
lead to any spectacular events
like formation of a large hole in the ground. However, as it was argued in
\cite{b24} -- \cite{b31}, baby universe
formation may  lead to important modifications of the properties of our vacuum
state.

One should note also, that in the eternally existing self-reproducing
inflationary
universe this process may occur even at the present time in those domains of
the
universe which are now in the inflationary phase at a density close to the
Planck
density \cite{b19,MyBook}. Quantum jumps of the scalar field to the Planck
density space-time foam, which
regularly occur in this scenario, and subsequent jumps of the field back from
the
space-time foam may be interpreted as a process of creation of new inflationary
universes which are not attached to our universe by any regions of classical
space-time.
Of course, one may argue  that the distance from us to the inflationary domains
with the Planck density
 is exponentially large. Moreover, these regions may form huge black holes and
become effectively disconnected from our part of the universe \cite{b32}. One
should remember, however, that in the end we are going to investigate a {\it
non-local} interaction of baby universes with our universe.  It is important
that
 {\it the main part} of the physical volume of a self-reproducing inflationary
universe (in the
synchronous coordinate system) is always occupied by the fluctuating scalar
field with $V(\phi) \sim M_p^4$ \cite{b19,MyBook}. Moreover, it is clear that
if
the process of the baby universe formation may occur at all, it should be most
efficient at the density close to the Planck density. This suggests (see also
\cite{Clain,Mezh})  that an investigation of the baby universe formation during
inflation may be very important for the understanding of  the properties of the
gravitational vacuum.

The possibility that our own universe could appear as a result of decoupling
from another universe may have other interesting implications, some of which
may
be  experimentally testable. For example, if our universe was created ``from
nothing'', then it typically appears in a state with a very large vacuum energy
density, $V(\phi) \sim M^4_p$ \cite{b56}, and later it enters an infinite
process of self-reproduction \cite{b19}.  In such case it does not seem
possible
to avoid the standard prediction of inflationary cosmology, $\Omega = 1$.
However, if our universe is created due to decoupling from another universe,
then from our estimates  of $P(\phi)$ it follows that in some  theories the
universe should be typically created  in a state with a small value of
$V(\phi)$, which only leads to a very short stage of inflation, if
any.\footnote{This issue is not very trivial, since the expectation value of
the
{\it volume} of the inflationary universe in the theories with $V(\phi) \sim
\phi^n$ (without an account taken of its self-reproduction at very large
$\phi$)
is proportional to $P(\phi) \exp\left({12\pi \phi^2\over n M_p^2}\right)$
\cite{MyBook}, which may grow at large $\phi$ even if $P(\phi)$ decreases.} In
such case our universe may be relatively small, and it may have $\Omega \not =
1$.  We do not think that this model is natural, but it is better to know that
such a possibility may exist. We mentioned it here, since it illustrates a
novel
way to solve the homogeneity and isotropy problem without much help of
inflation:
 It is well known that only spherically symmetric bubbles are formed if the
probability of their creation is strongly suppressed.

As for the possibility to create the universe in a laboratory, our estimates
indicate that one would need a very good laboratory indeed.
The probability to create the universe in a laboratory is not totally
negligible
only in the chaotic inflation scenario, only in the theories with large
coupling
constants or with scalar particles with masses of the order of
$M_p$, and only if one can heat  the system up to the temperature approaching
$M_p$.
However, the most ironical part of    it is the question whether the new
universe can be useful for us in any way.  Of
course, one may just consider the problem of the universe creation  as an
interesting theoretical problem to think about in a spare time, but if the
universe creation is entirely useless, one may find other interesting
 problems to solve. Leaving aside the possibility to use the universe as a
universal trash compactor,
we were hardly able to find any good reason to spend our time and energy for
its
creation.

Indeed, one cannot ``pump'' energy  from the new universe to ours, since this
would contradict the energy
conservation law. One cannot jump into the new universe, since  at the moment
of
its creation it  is
microscopically small and extremely dense, and later it decouples from our
universe. One even cannot send any
information about himself to those people who will live in the new universe. If
one tries, so to say,
to write down something ``on the surface of the universe'', then, for the
billions of billions years to come, the
inhabitants of the new universe will live in a corner of one letter. This is a
consequence of a general rule: All
local properties of the universe after inflation do not depend on initial
conditions at the moment of its
formation.  Very soon it becomes absolutely flat, homogeneous and isotropic,
and
any original message
``imprinted'' on the universe becomes unreadable.

We were able to find only one exception to this rule.  As we already mentioned,
if chaotic inflation starts at a sufficiently large energy density, then it
goes
forever, creating new and new inflationary domains. These domains contain
matter
in all possible ``phase states'' (or vacuum states),  corresponding to all
possible minima of the effective potential and all types of compactification
compatible with inflation \cite{b19,MyBook}. However, if inflation starts at a
sufficiently low energy density,  as is often the case with the universes
produced in a laboratory, then no such diversification occurs; inflation at a
relatively small energy density does not change the symmetry breaking pattern
of
the theory and the way of compactification of space-time. Therefore it seems
that the only way to send a message  to those
who will live in the universe we are planning to create is to encrypt it into
the properties of the
vacuum state of the new universe, i.e. to the laws of the low-energy physics.
Hopefully, one may achieve it by choosing a proper  combination of temperature,
pressure and external fields,  which would lead to creation of the universe in
a
desirable phase state.

The corresponding  message can be long and informative enough only if there
are extremely many ways of symmetry breaking and/or patterns of
compactification
in the
underlying theory. This is exactly the case, e.g.,  in the superstring theory,
which was considered
for a long time as one of the main problems of this theory.
Another requirement to the informative message is that it should not be too
simple. If, for example,  masses of all particles would be equal to each other,
all coupling constants would be given by $1$, etc., the corresponding message
would be too short. Perhaps, one may say quite a lot by creating a  universe in
a strange vacuum state  with $m_p \sim 2000 m_e$, $m_W \sim 100 m_p$, $m_X \sim
10^{13} m_W$, $M_p \sim 10^4 m_X$. The stronger is the symmetry breaking, the
more ``unnatural'' are relations between parameters of the theory after it, the
more
information the message may contain. Is it the reason why we must work so hard
to understand
strange features of our beautiful and imperfect world? Does this mean that our
universe was created not by a
divine design but by a physicist hacker? If it is true, then our results
indicate that he did a very difficult job.
Hopefully, he did not make too many mistakes...

\section{Acknowledgements}

It is a pleasure to thank T. Banks, S. Coleman, A. Guth, J. Ellis, S. Hawking,
M. Miji\'c, M. Sher and  L. Susskind
for valuable discussions at different stages of this investigation. This work
was supported in part by NSF grant
PHY-8612280.

\end{document}